\documentstyle[aps]{revtex}

\makeatletter
\def\thepage{3-\@arabic\c@page}
\def\@pnumwidth{2em}
\def\REVTeX{REV\TeX}
\makeatother

\newcommand{\mgsio}{MgSiO$_3$}

\newcommand{\ka}{\kappa}
\newcommand{\phim}[6]{\Phi_{#1 #2}\left( ^{#3}_{#5} \ ^{#4}_{#6} \right)}
\newcommand{\xind}[2]{ \left( ^{#1} _{#2} \right) }
\newcommand{\bphi}{${\bf \Phi}$}
\newcommand{\bd}{${\bf D(q)}$}
\newcommand{\castep}{{\sc castep}}

\begin{document}

\title{
Ab initio studies of structural instabilities in magnesium silicate perovskite
}

\author{M.C. Warren and G.J. Ackland}
\address{Department of Physics and Astronomy, The University of Edinburgh,
Mayfield Road, Edinburgh, EH9 3JZ, Scotland, UK}

\maketitle

\makeatletter
\global\@specialpagefalse
\def\@oddhead{\REVTeX{} 3.0\hfill Released November 10, 1992}
\let\@evenhead\@oddhead
\def\@oddfoot{\reset@font\rm\hfill \thepage\hfill
\ifnum\c@page=1
  \llap{\protect\copyright{} 1992
  American Institute of Physics}
\fi
} \let\@evenfoot\@oddfoot
\makeatother

\begin{abstract}

Density-functional simulations are used to calculate structural
properties and high-symmetry phonons of the hypothetical cubic phase,
the stable orthorhombic phase and an intermediate tetragonal phase of
magnesium silicate perovskite.  We show that the structure of the stable
phase is well described by freezing in a small number of unstable
phonons into the cubic phase.  We use the frequencies of these unstable
modes to estimate transition temperatures for
cubic$\rightarrow$tetragonal and tetragonal$\rightarrow$orthorhombic
phase transitions.  These are investigated further to find that the
coupling with the strain suggests that phonons give a better
representation than rigid unit modes.  The phonons of an intermediate
tetragonal phase were found to be stable except for two rotational
modes.  The eigenvectors of the most unstable mode of each of the cubic
and tetragonal phases account for all the positional parameters of the
orthorhombic phase.  The phase boundary for the orthorhombic--tetragonal
transition intersects possible mantle geotherms, suggesting that the
tetragonal phase may be present in the lower mantle.

 \end{abstract}

\section{Introduction}

Magnesium silicate makes up most of the material in the Earth's lower
mantle, and its properties and phase transitions determine much of the
density distribution, thermal properties and discontinuities of the
mantle (Catlow and Price 1990, Navrotsky and Weidner 1989).  Such
discontinuities, which may be chemical (involving substitution of iron
for magnesium) or physical transitions, have been observed in the lower
mantle at several depths (Stixrude and Cohen 1993, Kawakatsu and Niu,
1994).  Convection in the upper and lower mantles, caused by
temperature gradients, is also a topic under debate (Bukowinski
and Wolf 1988, Jeanloz and Morris 1986), and the possibility of
temperature-induced phase transitions in \mgsio\ may help to ascertain
the actual behaviour of the mantle.

The phase of \mgsio\ believed to exist in the mantle has the perovskite
structure displayed by many other compounds.  This structure exhibits a
wide variety of behaviour, with a multitude of possible phase
transitions (Lines and Glass 1977).  Despite much study (reviewed by
Navrotsky and Weidner (1989) and Hemley and Cohen (1992)), the
properties of \mgsio\ are still not fully understood: attempts to
produce samples for X-ray diffraction under mantle pressures have met
with difficulty (Hemley et al. 1987, Mao et al. 1991, Ross and Hazen
1990).  However, there has been evidence for an
orthorhombic--tetragonal--cubic series of transitions from observations
of twinning in the orthorhombic phase after quenching from high
temperatures (Wang et al. 1992).

This work considers the observed orthorhombic ({\it Pbnm}) phase, the
hypothetical high-temperature cubic phase ({\it Pm3m}) and an
intermediate tetragonal phase ({\it I4/mcm}); symmetry increases with
increasing temperature.  The vibrational dynamics of these modes is
very important since transitions between these phases are likely to
proceed via soft-mode mechanisms (Hemley and Cohen 1992).  We calculate
the frequencies and eigenvectors of some high-symmetry phonons of all
three phases using density-functional simulations, and also investigate
which phonons would be involved in transitions between the three
phases.

The triply degenerate zone-centre phonons of several other perovskites
have already been found from first principles (Cohen and Krakauer 1990,
Postnikov et al. 1994 and King-Smith and Vanderbilt 1994).  The
eigenvectors calculated by King-Smith and Vanderbilt (1994) correctly
predict the symmetry of distorted phases with one formula unit in the
unit cell (i.e.  those in which only zone-centre phonons can freeze
in).  However, the orthorhombic phase of \mgsio\ contains twenty atoms,
as opposed to the five atoms in the cubic unit cells, so at least some
of the phonons involved in any transition between cubic and
orthorhombic phases must lie away from the zone centre of the cubic
phase.  A simulation of more than one unit cell of the cubic crystal is
thus required to calculate these zone-boundary phonons.

The transition temperatures required for the cubic and tetragonal phases
were estimated by using the phonon frequencies to parameterise a simple
model.  The variation of these temperatures under pressure is deduced
from a combination of first-principles Gruneisen parameters and bulk
modulus, and experimental thermal expansivities; compensation for the
effect of the Local Density Approximation (LDA) is discussed in
Section~\ref{sect:discuss.lda}.

The structure of this paper is as follows: Section II gives the phases
studied; in Section III we describe our {\it ab initio} methods;
Sections IV, V and VI give results for cubic, tetragonal and
orthorhombic phases respectively; Section VII analyses the important
phonons of the cubic phase and considers the role of the tetragonal
phase.  Section VIII discusses calculations of transition
temperatures.  Sections IX and X discuss the conclusions which have been
reached.

\section{Structure}

All phases of \mgsio\ perovskite essentially consist of octahedral
`cages' of oxygen atoms, linked at the corners, each containing a
silicon ion in the centre.  In the spaces between the octahedra there
are 12-fold coordinated magnesium ions.  However, in the
orthorhombic structure, the octahedra are rotated around the silicon
(B) ions, and the magnesium (A) ions are displaced, as shown in
Figure~\ref{fig:str.ortho}, giving space group {\it Pbnm}.  This is the
form of \mgsio\ observed under ambient conditions (Yagi et al. 1978) and
assumed to be the dominant phase in the lower mantle.  The tetrahedral
structure is an intermediate, having octahedral rotations about only one
axis, with {\it I4/mcm} symmetry; the reasons for its consideration are
given in Section~\ref{sect:form.tetrag}.

The oxygen octahedra are nearly perfectly rigid.  The `rigid unit' model
of such structures (Giddy et al. 1993) assumes that they are
incompressible and cannot be distorted.  If they rotate, continuing to
share oxygen atoms at the corners, the unit cell vectors perpendicular
to the rotation axis must therefore decrease in length.  Such rotational
modes are known as `rigid unit modes'.  At the other extreme, true
phonon modes must preserve the size and shape of the unit cell, so the
octahedra must enlarge and distort slightly if they are to rotate.  The
applicability of these models to \mgsio\ is investigated and discussed
in more detail in Section~\ref{sect:rums}.

\section{Ab initio lattice dynamics}

Density functional simulations have already been used to investigate
lattice dynamics in many materials, e.g. Srivastava and Kunc (1988).  The
most popular `direct' approach involves finding the change in energy
and/or forces when one atom is displaced from its equilibrium position
by a small amount, to give some of the harmonic force constants \bphi\
of the crystal.  A crystal is divided into $L$ primitive cells, labelled
by $l$, each containing $r$ atoms, labelled by $\ka$.  If the
displacement from equilibrium of the atoms is given by ${\bf
u}\xind{l}{\ka}$ and the forces on the atoms by ${\bf F} \xind{l}{\ka}$,
the force constants are defined by

\begin{equation}
F_{\alpha} \xind{l}{\ka} =
\sum_{l'} \sum_{\ka'} \sum_\beta \phim{\alpha}{\beta}{l}{l'}{\ka}{\ka'}
     u_{\beta} \xind{l'}{\ka'}
\end{equation}

The periodicity of the crystal, and hence of any phonon, can be used to
reduce the problem of solving $3Lr$ equations of motion to $3r$
(Maradudin et al. 1971).  The frequencies $\omega$ and eigenvectors
{\boldmath $\epsilon$} of the phonons at a wavevector ${\bf q}$ are now
found by solving the eigenproblem:

\begin{equation}
-\omega^2 \mbox{\boldmath $\epsilon$} = {\bf D(q)} \mbox{\boldmath $\epsilon$ }
\label{eqn:neat.form}
\end{equation}

The Fourier-transformed dynamical matrix \bd is given in mass-reduced
coordinates by

\begin{equation}
D_{\alpha \beta}(\ka \ka'|{\bf q})
             =  \frac{1}{\sqrt{m_\ka m_{\ka'}}}
\sum_{l}\phim{\alpha}{\beta}{0}{l}{\ka}{\ka'}
               \exp(i{\bf q}.[{\bf x}\xind{l}{\ka'}-{\bf x}\xind{0}{\ka}])
\label{eqn:define.D}
\end{equation}

Hence knowledge of all the force constants \bphi\ of the structure
enables all the phonons to be found.  As it stands, the phonons at any
wavevector can be found, but all the force constants of the infinite
crystal must be known.  A simulation of a `supercell' consisting of a
small number of unit cells can give only specific combinations of force
constants.  However, periodic boundary conditions restrict the phonons
which may exist in such a supercell to those which have the periodicity
of the supercell.  For these modes, the combinations of force constants
can be treated as the true force constants if the sum over $l$ in
(\ref{eqn:define.D}) now runs over only the unit cells in the supercell.
In general, this restriction applies when the range of the forces is
greater than the size of the supercell.  Investigations in this case
show this to be so, probably due to the large degree of ionicity in the
structure.  This method of calculating phonons thus does have
limitations, but is simple to apply to existing density-functional
schemes; in our case we are only concerned with phonons at high-symmetry
points of the Brillouin Zone, which can be obtained in this way.

We used a self-consistent density-functional scheme within the LDA,
employing norm-conserving, nonlocal, Kleinman-Bylander pseudopotentials
(Kleinman and Bylander 1982, Kerker 1980).  A plane-wave basis set was
used for the electronic wavefunctions, with energies up to a cutoff of
${\cal E}_{\mbox{\small cutoff}}=$500 eV, corresponding to
approximately 4000 plane waves at each $k$-point for a twenty atom
cell.  A Monkhorst-Pack $k$-point scheme (Monkhorst and Pack 1976) was
used to give 4 $k$-points in the Brillouin Zone.

In this work, an adaptation of the \castep\ code was used
(Payne et al. 1992) which gives the Hellmann-Feynman forces on each
atom, from which the force constants can be calculated.  Each simulation
of the crystal with one atom displaced from equilibrium gives one row of
\bphi.  In a structure with high symmetry, only a few such simulations
are needed to construct the entire dynamical matrix, by employing
symmetry elements of the space group of the crystal.  If an element $\{
{\bf S} | {\bf v}(S) \}$ maps atom $j$ to $J$ and atom $\ka$ to $K$ then
it can be shown (Maradudin et al. 1971, pages 10--14) that a pair of
force constants are related by

\begin{equation}
{\bf \Phi} (JK) = {\bf S \Phi} (jk) {\bf S}^T
\end{equation}

Hence if all the symmetry elements of the structure are known, all the
force constants can be deduced from a minimum number of simulations.

\section{Cubic phase}

The high-symmetry cubic phase has a unit cell of only one formula unit
(5 atoms) and is thus the fastest to simulate.  It shares many features
with the tetragonal and orthorhombic phases, so much of this discussion
applies to all three phases.

\subsection{Equilibrium structure}

The equilibrium lattice parameter was found by a first-principles
quenched molecular dynamics simulation of a cubic unit cell of 5 atoms,
allowing the volume $\Omega$ of the simulation supercell to change under
the stress, which was also calculated from first principles.  A
modification of the fictitious Parrinello-Rahman Lagrangian was used,
which gives the supercell a `boxmass' $W$ (Parrinello and Rahman 1980,
Wentzcovitch 1991).  A matrix $h$ is formed from the supercell vectors
$\{ {\bf a,b,c} \}$, giving a metric tensor $g=h^T h$.  The fractional
coordinates of the $N$ atoms are denoted by ${\bf s}_i$ with masses
$m_i$, so ${\bf r}_i = h{\bf s}_i$.  Another tensor is created from the
initial cell faces: $f_0 = (\sigma_0^T \sigma_0)$, where $\sigma_0 = \{
{\bf b}_0\! \times \! {\bf c}_0, {\bf c}_0 \! \times \! {\bf a}_0, {\bf
a}_0 \!\times \!{\bf b}_0 \}$ (Wentzcovitch 1991) which helps to
eliminate spurious rotation of the cell.  The Lagrangian is now

\begin{equation}
{\cal L} = \sum_{i=1}^N \frac{m_i}{2}{\bf s}_i^T \! \cdot g {\bf s}_i
- U({\bf r}_i) - P \Omega + \frac{W}{2} \mbox{Tr}(\dot{h} f_0 \dot{h}^T)
\end{equation}

However, if the cell changes then the plane wave basis set also changes,
and since it is computationally convenient to keep the number of plane
waves fixed, this means that the energy of the most energetic plane wave
will change.  However, it is physically more sensible to keep the cutoff
energy constant (Dacosta et al. 1986), and so a Pulay correction was
added to the stress (Francis and Payne 1990) to compensate:

\[
\Pi_P = \frac{2}{3 \Omega}
\frac{\partial E}{\partial (\ln {\cal E_{\mbox{\small cutoff}}})}
\]

The total energy and stresses were thus those which would be obtained if
the cutoff energy were kept constant at 500eV.  The structure stabilised
at a lattice parameter of $a_0=3.44$~\AA; this is consistent with other
calculations to within the effect of the LDA.  Further comment will be
made on this effect below.  The ionic positions are fixed by the cubic
symmetry, which was enforced throughout the simulation.  Approximately
twenty PR steps were required to obtain the equilibrium cell.

This method was checked by performing simulations at different cell
sizes, to find the bulk modulus.  This confirmed a minimum energy at
$a_0=3.44$~\AA.  The bulk modulus $K$ is sensitive to the fitting
procedure: a quartic polynomial fit (R=0.99992) gave $K$=255~GPa,
$K'=\partial K/\partial P=4 \pm 1$, in agreement with
Wentzcovitch et al. (1993) and a Murnaghan fit gave $K=241\pm3$~GPa,
$K'=3.6\pm0.1$.  These high results are comparable with the experimental
values for the orthorhombic phase: $K=266\pm6$~GPa and
$K'=3.9\pm0.4$; these are higher than most other oxide perovskites
(Navrotsky and Weidner 1989).

The bonding was largely ionic, as shown by the plot of charge density
in Figure~\ref{fig:charge}.  This is in agreement with the findings
reported in Navrotsky and Weidner (1989), and confirms that long-range
forces are important.

\subsection{Phonons}

\subsubsection{Zone centre}

A simulation of only one primitive cell (5 atoms) can only give
reliable information about the fifteen zone centre phonons, due to the
use of periodic boundary conditions over a distance smaller than the
range of the forces.  Four displacements were needed to construct the
zone-centre dynamical matrix: one each for the Si and Mg, and two for O
(parallel and perpendicular to cell face).  A separate simulation was
performed for each distortion, using the appropriate reduced symmetry.
The entire dynamical matrix was constructed for one unit cell, with the
help of the full symmetry of the cubic structure.  This matrix was
diagonalised to find five sets of triply degenerate $\Gamma$ phonons;
because the phonons have wave vector strictly equal to zero, these are TO and
TA modes (Cohen and Krakauer 1990).

The eigenvectors of the $\Gamma$-point modes are shown in
Table~\ref{table:emodes.cubic}, in mass-reduced coordinates
$\mbox{\boldmath $\epsilon$}_\ka = \sqrt{m_\ka}{\bf u}_\ka$.  It is
noteworthy that the unstable mode consists of motion of the magnesium
(A) ions against the rest of the crystal, in contrast to most
perovskites, in which it is the B atoms which `rattle' inside the oxygen
cage (King-Smith and Vanderbilt 1994, Lines and Glass 1977).  We find no
evidence of strong covalent Si--O bonds in \mgsio, so this motion is
probably due to the relative sizes of the oxygen, silicon and magnesium
ions (Hemley et al. 1987).

Three acoustic translation modes are expected with zero frequency; this
requires each row of \bphi\ (corresponding to the forces from the
displacement of one coordinate) to sum to zero (Maradudin et al. 1971).
This was achieved by adjusting the diagonal (self-interaction) elements
of the dynamical matrix by less than 1\%, giving an indication of the
accuracy of the calculations.  Anharmonicity was investigated by doing
a similar set of runs with twice the displacement, giving force
constants which differed from the originals by less than 0.1\%, so
anharmonicity was considered negligible.  The effect of $k$-point
sampling on phonons of other perovskites was found to be significant in
(King-Smith and Vanderbilt 1994).  However, we increased the $k$-point
set from 4 points in the Brillouin Zone to 32 and found that the
changes in frequency were approximately 0.2--0.4~THz, and that the
changes in eigenvectors 10\% at most.  The effect of $k$-point sampling
is even less in simulation of larger cells.

\subsubsection{Zone boundary}

A larger simulation is needed to investigate phonons away from the zone
centre.  A supercell of four primitive unit cells ($a'=b'=\sqrt{2}a_0$;
$c'=2a_0$; 20 atoms) was chosen, since this reflects the geometry of the
orthorhombic phase.  The lattice parameter $a_0$ obtained above was
used, and the force constants obtained in a similar manner to the 5 atom
cell.

Because the supercell was not cubic, displacements in the $z$ direction
as well as the $x$ direction were needed.  The Fourier-transformed
dynamical matrix \bd\ was then constructed at the zone centre and
boundaries.  The frequencies obtained are shown in
Figure~\ref{ortho.phonons}.  The shape of the supercell also had the
effect of breaking the cubic symmetry so that phonons are not perfectly
degenerate, e.g.  at the $R$ point of the Brillouin Zone.

Unstable modes can be described by a simple model consisting of quartic
local potentials interacting harmonically (described more fully in
Section~\ref{sect:spring.model}).  These give complete unstable bands
if the local potential is a double well type, giving order-disorder
type transitions (Bruce 1980), but if the local potentials have a
single central minimum (displacive type) part of the band will be
stable (Samuelsen et al. 1971, Zhong et al. 1994).  In this case, it was
found that there are unstable phonon modes at all points investigated,
suggesting that any phase transition involving this band would have
some order-disorder character.

The eigenvectors of all these modes were also obtained.  The most
unstable modes were found at $R$ and $M\!$, consisting of rotations of
near-rigid octahedra about the silicon atoms.  This is consistent with
the unstable modes found by Stixrude and Cohen (1993), Bukowinski and
Wolf (1988) and Hemley et al. (1987), but those workers found no other
unstable phonons, whereas we find several involving magnesium
displacement.  At $M\!$, the mode involves rotation about $z$, and is
denoted $M_2$; rotations at $R$ around all three axes are denoted
$R_{25}$.  These modes have been previously predicted to be the only
zone-boundary rigid unit modes in cubic perovskite (Giddy et al. 1993).
The stable modes mostly involve squashing and distorting the octahedra
and displacement of magnesium ions.

\subsubsection{Effect of pressure on frequency}

The CPU time required to simulate a fixed configuration of $N$ atoms
varies as $N^3\!$, so the five-atom cell required much less time than the
twenty-atom cell: the Hellmann-Feynman forces from one distortion could
be found in under half an hour on an Alpha AXP supercomputer.  It was
hence practicable to investigate the variation of $\Gamma$ mode
frequencies with cell size and obtain estimates of the Gr\"{u}neisen
parameters $\gamma_i = -\partial(\ln \omega_i)/ \partial \ln V$.  These
are important for predicting the thermal properties of materials.  The
results are shown in Figure \ref{fig:grunei}.  We found that larger
cells were more unstable with respect to the magnesium unstable mode,
and that the frequencies of all other $\Gamma$ phonons decreased as the
cell was enlarged, i.e.  have positive $\gamma$, with similar magnitudes
to those obtained experimentally in the orthorhombic phase
(Navrotsky and Weidner 1989).

The effect of pressure on the $M_2$ unstable mode was also investigated,
by rotating the octahedra at several different cell volumes.  It was
found that in this case the magnitude of the imaginary frequency
increased with pressure, i.e.  under compression the structure becomes
more unstable with respect to this distortion, in contrast to the
behaviour of the unstable mode at $\Gamma$.  This agrees with the trends
found by Hemley et al. (1987) and implies that pressure does not favour
the cubic phase.

\section{Phonons of intermediate structures}

There are two important intermediate tetragonal structures, with either
the $R_{25}$ or $M_2$ mode frozen into the cubic phase, to their local
equilibrium.  However, these structures may still contain some unstable
phonons corresponding to those previously unstable modes which are not
frozen in.  In both cases, we calculated the normal modes using a similar
procedure to that for the cubic phase.

\subsection{Freezing in $M_2$ mode}

We investigated the configuration with a $M_2$-rotation of the oxygen
octahedra frozen into the cubic phase, giving a unit cell of 10 atoms
with space group {\it I4/mmm}.  Only modes containing distortions in the
$xy$ plane were considered, since these $z$ distortions form a separate
space, and are likely to be very similar to those in the undistorted
cubic phase.

Phonons with the same symmetry as the frozen mode, i.e.  involving only
O{\sc ii} distortion, remained normal modes, although their frequencies
were changed.  Other normal modes are combinations of the original cubic
$\Gamma$ and $M$ phonons.

The $M_2$ rotation mode, which had a frequency of 11.1$i$ THz in the
cubic structure, is now stable with a frequency of 13.5 THz.  In a
quartic double-well model (see Section~\ref{sect:double.well}), the
stable frequency would be $\sqrt{2} \omega_0$ where $i\omega_0$ is the
unstable frequency.  The stable frequency found here is a little smaller
than this, showing a small departure from a pure quartic potential; this
is evident in the fit to a quartic in Figure~\ref{fig.coupl.mstrain}.
This phonon stiffened further in the full orthorhombic structure, to
14.1 THz.

The mode with the highest frequency in the cubic structure was still the
highest, but the frequency was reduced from 29.6 THz to 24.3 THz.  Two
degenerate unstable modes were found, predominantly consisting of
displacements of the magnesium ions, with the degree of instability
(3.66$i$ THz) slightly decreased from that in the cubic structure
(4.76$i$ THz).

\subsection{Freezing in $R_{25}$ mode}

\label{sect:form.tetrag}

The cubic $R_{25}$ mode can be frozen into the cubic supercell, to form
a twenty atom body-centred tetragonal cell (space group {\it I4/mcm}).
The cubic $R_{25}$ mode is degenerate; we chose rotation about the $x$
axis of the twenty-atom cell, since this is the rotation observed
in the orthorhombic phase.  At large amplitudes of $R_{25}$, the cubic
eigenvectors are no longer valid: non-negligible forces are now also
exerted on the magnesiums.  The general $R_{25}$ mode can therefore be
thought of as including some magnesium motion.  To obtain the
equilibrium amplitude, the structure was relaxed with the appropriate
symmetry; the magnesium moved along $x$ with approximately 25\% of the
O{\sc i} displacement.

Sixty phonons (at tetragonal $\Gamma$ and $X$) of this phase were found.
The eigenvectors were compared to those of the cubic cell phonons, and
each tetragonal phonon assigned to its closest match in the cubic
structure; in what follows we shall denote the phonons by their
equivalent cubic symmetry label.  The tetragonal and orthorhombic
phonons are plotted against this part of the cubic Brillouin Zone in
Figure~\ref{ortho.phonons} to enable comparisons to be made.  There are
only two unstable modes in this structure: the $R$ and $M$ modes
corresponding to rotation about the $z$ axis, with the latter slightly
more unstable (5.64$i$ THz).  The eigenvector of the $M$ mode consists
of $M_2$-type octahedral rotation together with some motion of Mg and
O{\sc i}.  We conclude that this is related to the original cubic $M_2$
mode.  This mode is much less unstable that its cubic counterpart.  All
other modes, including the magnesium displacement modes, which were
unstable in the cubic structure, are now stable.  Since there is no
longer a complete band of unstable phonons, a soft-mode transition from
this tetragonal phase to the orthorhombic would have displacive
character, in contrast to a cubic--tetragonal transition.

The $R_{25}$ mode is the most unstable in the cubic phase, and this
tetragonal structure has lower energy than the {\it I4/mmm} phase
discussed above, so we chose this structure as our intermediate
tetragonal phase between cubic and orthorhombic; on cooling from the
cubic phase, it will freeze in first.  The structure is shown in
Figure~\ref{fig:str.ortho}, and in the remainder of this paper we
consider the possibility that it forms a distinct thermodynamic phase.

\section{Orthorhombic phase}

\subsection{Structure}

The orthorhombic phase has much less symmetry than the cubic phase, and
hence requires more structural relaxation to find the equilibrium
structure.  As in the cubic phase, the three cell edges were allowed to
move under the internal and Pulay stresses, but the ions were also
relaxed under the Hellmann-Feynman forces, until the equilibrium
structure was obtained.  This simulation was started from values given
by Wentzcovitch et al. (1993) and the symmetry was enforced throughout the
run.  There are ten structural parameters, and so this relaxation took
a long time: several hundred hours of CPU time on a Alpha AXP.
The ionic motion was controlled by a conjugate-gradient routine, and
relaxed slowly compared to the cell.  The conjugate-gradient scheme is
not guaranteed to work for non-linear problems, such as ionic positions
coupled to each other and to the cell parameters. Near equilibrium, the
movement of the cell had to be terminated so that the structural
parameters could be found accurately.

\label{sect:discuss.lda}

The structural parameters obtained are given in
Table~\ref{table:struc.ortho}.  The unit cell is smaller than most
previous work, but the structural parameters follow the reported trend
of an increase in distortion under compression, (Hemley et al. 1987,
Wentzcovitch et al. 1993, Matsui 1988).  Following Bukowinski and Wolf
(1988) we compare the rotation angles $\theta=\cos^{-1}(a/b)$ of $M_2$
and $\phi=\cos^{-1}(\sqrt{2}a/c)$ of the $R_{25}$ rotations in
Table~\ref{table:rot.angles}, effectively assuming that the rotations
are rigid unit modes.  This confirms that our structure is the most
distorted.  However, the Si--O bondlength is only 1\% smaller than that
observed experimentally, which is typical for calculations using the
LDA (Payne et al. 1992).  These results suggest that the effect of LDA
can be approximated by an external pressure, since both volume and
structure are affected.  The effect of the LDA has also been compared
to a pressure by Zhong et al. (1994) and Akbarzadeh et al. (1993), and
generalised-gradient corrections have been similarly treated by
Kresse et al. (1994).

We can use this approach in comparing our work to previous
density-functional calculations of \mgsio: the cell parameters of
Wentzcovitch et al. (1993), when compared to the results of Ross and
Hazen (1990), suggest they are equivalent to a pressure of 8--10~GPa.
The effect of pressure on cell parameters calculated by
Wentzcovitch et al. (1993) suggests that the cell parameters of our work
correspond to an additional 10--12~GPa.  The experimental bulk modulus
and equilibrium volume give a pressure of 20~GPa for the volume found
in this work, which is consistent with these estimates.  Hence we
assume in what follows that our orthorhombic structure is under an
effective pressure of 20~GPa.

The effect of the LDA on structural parameters is harder to match to a
consistent pressure, since the experimental data in Ross and Hazen
(1990) does not show the simple monotonic trends of Wentzcovitch et al.
(1993) and does not extend to 20~GPa.  However, the experimental
data does show increasing octahedral rotation under pressure, and most
of our structural parameters continue the trends calculated by
Wentzcovitch etc al. (1993), so we conclude that the structural
parameters are also consistent with the LDA pressure approximation.

The remaining small differences between our calculations and
Wentzcovitch et al. (1993) are probably due to differences in pseudopotential,
which also cause the differing LDA pressures assumed.

To examine the coupling between distortion and strain, we have
also relaxed the ionic positions within the equilibrium cell of twenty
atoms of the cubic structure, i.e.  using $a=b=\sqrt{2}a_0$, $c=2a_0$.
In this simulation, the distortion from the cubic structure is about 5\%
less than that of the full orthorhombic cell found above.  This is
discussed in more detail in \ref{sect:rums}.  \label{sect:cub.relax}

\subsection{Orthorhombic phonons}

The symmetry of the orthorhombic phase means that displacements are
required along $x$, $y$ and $z$.  The ionic positions were relaxed to
give forces below 0.001 eV/\AA, so the remaining forces on the
`equilibrium' structure were subtracted from the forces on distorted
structures, assuming that the harmonic approximation applies to
sufficient accuracy.  To minimise anharmonic effects, the force
constants were averaged over pairs which must be symmetric; this process
changed the phonon frequencies by a maximum of 2\% for the optic phonons.

The highest frequency phonon was at 25.9 THz, which is much lower than
that of the highest cubic phonon.  As expected, there were no unstable
modes.  Only the $\Gamma$ phonons of the orthorhombic Brillouin Zone can
be found, since only one unit cell is simulated.  These correspond to
the $\Gamma$, $X$, $M$ and $R$ phonons of the cubic phase, due to the
quadrupling of the unit cell, and are again plotted in
Figure~\ref{ortho.phonons} against the appropriate part of the cubic BZ.

The phonon with the highest frequency at each point of the BZ is the
same in cubic, tetragonal and orthorhombic phases, although at lower
frequencies the orthorhombic eigenvectors are often linear combinations
of several cubic phonons.  The phonons which were unstable in the
cubic phase are now all stable, and mostly have low frequency.  The
eigenvectors are used to determine the symmetry of the phonons ($g$ or
$u$): the 24 symmetric modes were found, which are the Raman modes, and
within this group seven $A_g$ modes can be identified.  These are
likely to include the strongest Raman reflections, and are compared with
experimental values in Table~\ref{table:raman.modes}.

We further investigated the degree of anharmonicity by recalculating
some force constants using half the original displacement.  This only
had a small effect on the calculated phonons and eigenvectors: the
maximum change in frequency was 0.5~THz when force constants from Mg and
O{\sc i} $z$ displacements were recalculated in this way.

\section{Investigations of individual phonons}

\subsection{Cubic phonon modes contributing to orthorhombic structure}

We describe the structure of the stable orthorhombic phase by freezing
in phonons from the cubic phase.  Since these apply to displacements at
constant volume, we use only the fractional positions of the
orthorhombic cell, as if it had the same cell parameters as the cubic
cell, and allow coupling to the strain once these positions are fixed.
We define coefficients $c_j$ by

\begin{equation}
{\bf d} = \sum_j c_j {\bf p}^j
\label{eqn:decompose}
\end{equation}

where ${\bf p}$ are the mass-reduced eigenvectors of the cubic crystal,
and ${\bf d}$ is the mass-reduced displacement from cubic found in the
stable orthorhombic structure.  It is straightforward to find the
coefficients $c_j$ since the eigenvectors ${\bf p}$ are orthonormal,
and for the equilibrium orthorhombic cell we denote them $c_0$.

This is not an ideal description of the structure, since the harmonic
eigenvectors do not always describe the mode at large displacement.
However, the $\Gamma$, $X$, $M$ and $R$ cubic eigenvectors do span the
complete set of all possible strain-conserving distortions in a twenty
atom cell.  The orthorhombic cell consists of the equivalent of four
cubic unit cells, so we know that all distortions not involving strain
can be expressed by using cubic phonons as a basis set.
\label{sect:coeffs}

The results are shown in Table~\ref{tab.depths}.  Only two of the 42
stable phonons have significant non-zero coefficients, and only four of
the fifteen unstable cubic modes are required to form the orthorhombic
structure.  Phonons consisting mainly of magnesium displacement are
present much more weakly than those involving octahedral rotations, and
are very strongly coupled to the positions of the oxygens (discussed in
Section \ref{sect:form.tetrag}).  Hence we concentrate on the octahedral
rotation phonons.  We have also calculated these coefficients for the
distorted structure in the cubic cell, where we find all coefficients to
be smaller.

\subsection{Energy surfaces of rotational modes}

\label{sect:double.well}

A single unstable mode which becomes stable after a certain displacement
can be considered as motion in a potential of a double well form.  We
will take as a first approximation to the energy

\[
E(c) = -\alpha c^2 + \beta c^4
\]

where $\alpha$ and $\beta$ are positive constants and $c$ is the
amplitude of one phonon mode.  $E(c)$ has a central maximum and a
minimum on either side, as shown in Figure~\ref{fig:model.well}.  In the
case of the unstable modes in cubic perovskite, the cubic phase
corresponds to the maximum at $c=0$, but there are minima at $c' =
\sqrt{\alpha /2 \beta}$ which correspond to a stable distorted phase.
In the absence of phonon-phonon or phonon-strain coupling, the $c_0$
given in Table~\ref{tab.depths} would correspond to these minima, and we
have estimated the depths of the associated wells $E_0$ on this
assumption.

However, if two or more such phonons are present in a system, with
amplitudes $c_1$ and $c_2$, there will in general be a coupling between
them.  This could be represented by, for example:

\begin{equation}
E(c_1,c_2) = -\alpha_1 c_1^2 + \beta_1 c_1^4 - \alpha_2 c_2^2 +
\beta_2 c_2^4 + h c_1^2 c_2^2
\label{eqn.couple}
\end{equation}

The minima of this system are not at $(c_1',c_2')$ as defined above, but
depend on the coupling $h$.  For example, negative $h$ results in minima
which are deeper and further from the origin than those of the
one-phonon well.  The frequencies of the normal modes of the
one-dimensional model at the origin are unchanged by coupling, but at
the saddle points corresponding to the equilibrium amplitude $c'$ of one
mode, the instability of the other mode is increased by negative
coupling or decreased by positive coupling.

The coefficients obtained from the orthorhombic structure, using
Equation~(\ref{eqn:decompose}), are those of the overall minimum of all
phonon modes, not of isolated wells.  The orthorhombic structure
contains only distortions with $\Gamma$, $X$, $M$ and $R$ cubic
symmetry, as described in Section~\ref{sect:coeffs}.

To investigate this coupling in \mgsio, we can `freeze in' one cubic
phonon at a time into the cubic phase, and use the energies and
forces to parameterise the model described in (\ref{eqn.couple}).  We can
then compare the amplitude $c_0$ of each cubic phonon required to
describe the orthorhombic structure with the equilibrium amplitude $c'$
from such a simulation of the isolated cubic phonon.

The eigenvectors of the $M_2$ octahedral rotation remain unchanged as
the amplitude $c$ increases.  However, there can be coupling to other
modes when the phonon amplitude is beyond the harmonic limit: for
example, large $R_{25}$ displacement causes forces on the Mg atoms in
the $x$ direction, so this coordinate was also relaxed to equilibrium.
This demonstrates the coupling between rotational and magnesium modes
suggested above, which was accounted for in freezing in $R_{25}$ to form
the tetragonal phase.

Table~\ref{tab.frozen} shows the results of this process for the $M_2$
and $R_{25}$ phonons.  The $M_2$ rotation was also combined with a
change of the size and shape of the unit cell, such that the size of the
octahedra remained constant, to simulate a rigid unit mode rather than a
phonon.  The results are shown in Figure~\ref{fig.coupl.mstrain}, and
discussed further in Section~\ref{sect:rums}.

We find that for the $M_2$ rotation, the orthorhombic structure has a
smaller rotation angle than that of the isolated phonon in equilibrium.
 The equilibrium amplitude of the isolated $R_{25}$ rotation, including
coupling with Mg displacement modes, is smaller in the orthorhombic
structure than in the isolated phonon.  Thus the coupling $h$ between
these modes, defined in (\ref{eqn.couple}), is positive.  In both
cases, this effect is even stronger when the unit cell is not relaxed,
so this is a true phonon-phonon interaction rather than due to
interaction via the strain.  This is in agreement with similar
investigations using modified electron gas theory (Wolf and Bukowinski
1988) but differs from the negative $h$ value found by Stixrude and
Cohen (1993), where it is not clear whether coupling with Mg distortion
modes was considered.  It is also consistent with the reduction in the
instability of the $M_2$ mode when the $R_{25}$ mode is frozen in.

We used the values of $\omega_R/\omega_M$, $c'(M)$, $c_0(R)$, and
$c_0(M)$ to parameterise the model in Equation~(\ref{eqn.couple}), to
give $\beta_1=0.01683$, $\alpha_2=1.13$, $\beta_2=0.01150$ and
$h=0.0085$ when $\alpha_1$ is normalised to unity.  The amount of
coupling can be considered as $h^2/4\beta_1 \beta_2 \simeq 0.09$; the
differences between the equilibrium amplitudes of isolated and coupled
phonons are small.  When the $c_0$ values from relaxation in a cubic
cell were used, the coupling increased to $h^2/4 \beta_1 \beta_2 \simeq
0.15$, since the normal mode coefficients were even more reduced from
their isolated values.

\subsection{Coupling of rotational modes to strain}

\label{sect:rums}

The orthorhombic unit cell has a volume only 93\% of that of the cubic
phase, but the oxygen octahedra are larger by about 9\% in volume.  The
angle between Si--O{\sc i} and Si--O{\sc ii} bonds is 2$^\circ$ from
perpendicular, so there is also a small degree of distortion.  The
orthorhombic relaxation of ions within the cubic cell described in
Section~\ref{sect:cub.relax} (i.e.  allowing only pure phonon modes to
freeze in) gave octahedra 18\% larger than in the cubic phase, with
approximately 1$^\circ$ distortion.

This was further investigated as described above, by changing the unit
cell to be such that the size of the octahedra was constant while an
$M_2$ rotation was frozen in, to create a rigid unit mode.  The
resulting potential well had a minimum at approximately the same
rotation angle, but was not quite as deep, implying that this
combination is not as good a description as a phonon rotation with fixed
unit cell, in which the octahedra expand.  These results are shown in
Figure~\ref{fig.coupl.mstrain}.

The stress on the unit cell was also calculated (with Pulay corrections)
for the two cases: the phonon rotation with constant unit cell had
positive stress but the rigid unit mode with constant octahedra always
had negative stress.  This implies that the absolute minimum of rotation
and strain lies somewhere between these two extreme pictures.  This is
consistent with the equilibrium orthorhombic structure, in which the
unit cell and octahedral size are intermediate between phonons and rigid
unit modes.  In the case of relaxed ions in the cubic cell, the
enlargement of the octahedra is much greater, since there can only be
pure phonon modes.


\section{Transition temperatures}

The transition temperature between the orthorhombic and cubic phases
has been naively estimated by Stixrude and Cohen (1993) as $\Delta
E/k_B$, where $\Delta E$ is the energy difference between the two
phases for one cubic unit cell.  Applying this to our results gives
$T_c \simeq$ 20,000 K for the cubic--tetragonal transition, and
approximately 5,500 K for the tetragonal--orthorhombic transition.
These are slightly different from the results found by Stixrude and
Cohen (1993) which considered only octahedral rotations, and are both
considerably higher than the ambient temperature in the mantle.  These
estimates suggest that it is very unlikely that the orthorhombic--cubic
phase transition occurs in the mantle.

\label{sect:spring.model}

However, it is more reliable to calculate the transition temperature
from the energy stored in the interaction between unit cells, rather
than the total energy stored (Samuelsen et al. 1971).  The transition is
modelled as a one-dimensional quartic well (e.g.  local amplitude of
phonon eigenvector as coordinate) at each site, with harmonic
interactions between cells.  The lowest energy configuration consists of
these coordinates ordered with the wavevector ${\bf q}_c$ of an unstable
phonon, dependent on the form of the interaction.  The interaction
energy as a fraction of the total energy can be obtained from the
dispersion between $\Gamma$ and ${\bf q}_c$ of the appropriate phonon
band, as described in Appendix~\ref{appdx.tc}.  In this case we have two
modes contributing so we consider two such transitions.

We assume that a transition would proceed via the tetragonal
intermediate.  We can then consider the cubic$ \rightarrow $tetragonal
and tetragonal$ \rightarrow $orthorhombic steps separately.  For each
step, the strength of the interaction and local potentials were found
from the dispersion from $\Gamma$ to the appropriate part of the
Brillouin Zone (Appendix~\ref{appdx.tc}).  The fraction of the total
energy due to the interaction for the $R_{25}$ phonon, $(E_J/E_0)$, was
found to be 71\%, and 87\% for the $M_2$ phonon in the tetragonal phase,
the remainder being stored locally.  Using $k_B T_c = 2E_J/3$ from
Samuelsen et al. (1971), we find that the transition temperature for the
tetragonal--cubic step is approximately 10,000~K, which is still much
higher than typical mantle temperatures.  However, the orthorhombic to
tetragonal phase transition occurs at $T_c'=(E_{\mbox{\small
tet}}-E_{\mbox{\small orth}})/k_B \times E_J/E_0 \simeq$ 2,600~K, which
is quite possible in the mantle.  For the calculation of $E_0 =
E_{\mbox{\small tet}}-E_{\mbox{\small orth}}$, the unit cell of the
tetragonal phase was relaxed to equilibrium, as well as the structural
parameters, giving an orthorhombic structure with space group {\it
Imma}; this decreased the energy difference between this and the
orthorhombic phase.  For simplicity this intermediate phase is still
referred to as `tetragonal'.

Since the $M_2$ cubic phonon becomes more unstable with compression,
pressure will have a large effect on these transitions.  The
temperatures given above are for the ambient pressure in our
simulations, which is nominally zero pressure, and do not take account
of thermal expansion.  However, as discussed above, a unit cell which is
7\% too small and the observed structural parameters are consistent with
an external pressure of 20~GPa.  The effective pressure used in
calculations of the phase boundary is thus mantle pressure less this LDA
pressure.

We thus calculated the phase boundary using $k_B T_c' = \frac{2}{3} (E_0
+ P\Delta V)(E_J/E_0)$: $(E_J/E_0)$ is the fraction found above, varying
with pressure and temperature via Gruneisen parameters estimated from
the cubic phase and experimental thermal expansivity; $P$ is the
effective pressure; $\Delta V$ is the difference in volume between the
two phases at pressure $P$; $E_0$ is the energy difference between
equilibrium orthorhombic and tetragonal phases at zero temperature and
LDA pressure.  This phase boundary is shown in
Figure~\ref{fig:phase.bdy}, with a range of estimates of lower mantle
geotherms (Jeanloz and Morris 1986, Poirier 1991).

\section{Discussion}

We have found that only a small subset of the cubic phonons are
required to freeze in to create the orthorhombic phase.  Many unstable
phonons involve rotations of nearly-rigid oxygen octahedra, and it is
mainly these that are involved in forming the tetragonal and
orthorhombic phases from the cubic.  Some stable phonons are involved,
due to coupling between the unstable and stable modes away from the
harmonic limit.  This work finds that several modes predominantly
involving displacement of magnesium ions are also unstable in the cubic
phase, in contrast to previous studies (Bukowinski and Wolf 1988,
Hemley et al. 1987).  We find that the magnesium displacement observed
in the orthorhombic phase is due to coupling of magnesium modes into
the octahedral rotation modes at large amplitude, rather than a
separate freezing-in of magnesium modes.  Hence we have concentrated on
the octahedral rotation.  The cubic structure becomes more unstable
under compression, suggesting that the cubic structure is not favoured
by high pressure.

The coupling to the strain was found to be small: there was little
difference between phonons at constant volume and rigid unit modes, and
the fractional positions of relaxed ions in a cubic unit cell were very
similar to those with a relaxed unit cell.  Rigid unit models are known
not to be a perfect description of perovskite rotational modes (Giddy
et al. 1993) and this work suggests that phonons are in fact a better
description of these modes since they give lower total energy.  The
true picture lies somewhere between these two extremes, with octahedra
which are not perfectly rigid, as observed in the equilibrium
orthorhombic structure.  Pure rigid unit mode transitions also exhibit
displacive behaviour (Sollich et al. 1994), whereas the distribution of
unstable modes throughout the Brillouin Zone suggests that a transition
from the cubic phase would have some order-disorder character.

There is positive coupling between the two rotational unstable cubic
phonons frozen into the orthorhombic phase ($M_2$ and $R_{25}$), such
that the overall minimum is less distorted that individual distortions
would suggest.  This coupling is also manifested by the decrease in the
instability of the $M_2$ phonon after a $R_{25}$ phonon (the most
unstable in the cubic phase) is frozen in.  Because freezing in the
$R_{25}$ mode only partially stabilises the $M_2$ rotation, this forms
an intermediate tetragonal phase and the $M_2$ rotation is required to
freeze in to form the orthorhombic phase.  This tetragonal structure
may, however, exist as a distinct thermodynamic phase (Hemley and Cohen
1992) above some transition temperature $T'_c$, which will be lower
than the temperature $T_c$ required to obtain the cubic phase.

The tetragonal structure has displacements from the cubic which account
for three of the seven positional parameters of the orthorhombic phase.
The $M_2$-like phonon is the most unstable tetragonal mode, and contains
the remaining four displacements necessary to reach the orthorhombic
phase, so we conclude that the cubic phase can be considered to be
formed from the orthorhombic in two soft-mode steps, via this phase.
The relative magnitudes and directions of the displacements in the
tetragonal phase and of the tetragonal $M_2$ phonon are in approximate
agreement with those observed in the orthorhombic phase; the difference
can be ascribed to coupling between the modes, and volume differences
between the tetragonal and orthorhombic phases.  The
tetragonal--orthorhombic transition would probably have displacive
character, since the relevant phonon band in the tetragonal phase is not
unstable at every point in the Brillouin Zone.

The calculation of the transition temperature assumes that the
transition would proceed via this intermediate phase ({\it I4/mcm} or
{\it Imma}, both referred to as tetragonal).  It is clear that a
complete orthorhombic--cubic phase transition is out of the range of
mantle temperatures.  However, we find that the phase boundary for the
orthorhombic--tetragonal transition intersects some models of lower
mantle geotherms near the boundary with the upper mantle; this
transition might thus account for some of the discontinuities observed
in the lower mantle.  These results are also supportive of experimental
observations of twinning in \mgsio\ perovskite (Wang et al. 1992),
which suggested a transition temperature of 1900$\pm$200~K at 26~GPa
for an orthorhombic--cubic or orthorhombic--tetragonal phase
transition.  Typical mantle temperatures are considered to be in the
range 2,000--3,200~K with the upper bound given by the
experimentally-determined melting temperature.  These bounds encompass
two models of the mantle, giving `hot' and `cold' geotherms depending
on whether the upper and lower parts of the mantle convect together or
separately (Jeanloz and Morris 1986).  Hence the possibility of a
tetragonal phase suggested by this work has important consequences for
the behaviour of the mantle.

\section{Conclusion}

We have found the structure and an important set of phonons of cubic,
tetragonal and orthorhombic phases of \mgsio.  The unstable phonons of
the cubic structure are distributed throughout the Brillouin Zone, but
all become stable after the transition to the orthorhombic structure.
The structure and phonons of a tetragonal intermediate consisting of one
frozen-in rotation accounts for the distortions in the orthorhombic
phase.  Rotations of octahedra play the largest part in the transition,
and are the most unstable.  The coupling between these rotation phonons
and the strain implies that phonons are a better description of these
modes than the rigid-unit model.  It seems unlikely that the full
orthorhombic--cubic transition could occur in the mantle, since the
energy required is much larger than that available at usual mantle
temperatures.  However, a tetragonal intermediate may well be accessible
to mantle temperatures, and merits further study to see whether it might
explain some of the observed discontinuities and hence have consequences
for the properties of the whole mantle.  We believe that the sources of
error in our calculation are now smaller than the variation between
different mantle models.

\appendix{Calculation of $T_c$}

\label{appdx.tc}

A simple model for structural phase transitions uses one coordinate
$x_i$ at each site at position ${\bf r}_i$.  There is a quartic double
well at each site and harmonic inter-site coupling (the `$\phi^4$'
model described by Sollich et al. 1994), shown schematically in Figure
\ref{fig:model}.  The total static energy is written as

\[
U = \sum_i ( \alpha x_i^2 + \beta x_i^4) + \sum_{ij} J_{ij} x_i x_j
\]

The phonon frequencies of the undistorted phase at wavevector ${\bf q}$ are
then given by

\begin{eqnarray}
\omega^2({\bf q}) & = & 2 \alpha + J({\bf q}) \\
& & \mbox{where } J({\bf q}) = 2 \sum_j J_{0j} \exp i {\bf q} \cdot {\bf r}_j
\end{eqnarray}

If we assume nearest neighbour coupling (here between five-atom cells),
we have $J({\bf q})= -J(0)=J$, so we can find $\alpha$ and $J$ from the
calculated frequencies for the appropriate band.  The total energy
stored per site when all $x$ are totally ordered
(at $x_0^2=(\alpha+J/2)/2\beta$) with wavevector ${\bf q}$ is given by

\[
E_0 = \frac{(\alpha + J/2)^2}{4\beta} = \frac{x_0^2 (\alpha + J/2)}{2}
\]

and the total energy stored in the interaction per site is $J x_0^2 /
2$, so the fraction of the total energy which is stored in the
interaction is

\[
\frac{E_J}{E_0} = \frac{J}{\alpha + J/2}
\]

Since we know $E_0$ from {\it ab initio} total energy calculations, we
can hence find $E_J$, which is used to find the transition temperature.

\acknowledgments{

The authors thank M.C.~Payne for the original code and J.S.~Lin for
pseudopotentials.  M.C.W.  thanks the E.P.S.R.C.  for support and
S.J.~Clark, J.~Crain, G.S.~Pawley and V.~Heine for useful discussions;
G.J.A.  thanks B.P.  and The Royal Society of Edinburgh for a
fellowship, and the E.P.S.R.C.  for assistance in providing computing
facilities.

}

\begin{table}
\caption{Eigenvectors of the $\Gamma$ modes of cubic \mgsio, at $a=3.44$\AA.
Displacements are given in mass-reduced coordinates. An imaginary
frequency represents an unstable mode.}
\begin{tabular}{lllrrrrr}
  & $\omega$(THz) & Mg  & O{\small I} & O{\small II} & O{\small III}  & Si \\
\tableline
$\Gamma_{15}$ & 26.33 & 0.02   & -0.89  & 0.22      & 0.22       & 0.32  \\
$\Gamma_{15}$ & 13.37 & -0.253 & 0.070  & -0.398    & -0.398     & 0.783 \\
$\Gamma_{15}$ & 0.00  & 0.492  & 0.400  & 0.400     & 0.400      & 0.529 \\
$\Gamma_{15}$ & 6.01$i$ & -0.833 & 0.189 & 0.363    & 0.363      & 0.084 \\
$\Gamma_{25}$ & 5.189 &   0      &  0    &  -1      &  1         & 0 \\
\end{tabular}
\label{table:emodes.cubic}
\end{table}

\begin{table}
\caption{Structural parameters of equilibrium structure of orthorhombic
perovskite, with assumed pressure allowing for the effects of the LDA.
[1] Wentzcovitch et al. 1993; [2] Ross and Hazen 1990
}
\begin{tabular}{lrrrr}
       & this work & previous [1]
           & \multicolumn{2}{c}{experiment [2]} \\
         & 20~GPa & 10~GPa & 0~GPa & 10.6~GPa \\ \tableline
$a$(\AA) & 4.635  & 4.711  & 4.777 & 4.710 \\
$b$(\AA) & 4.833  & 4.880  & 4.927 & 4.873 \\
$c$(\AA) & 6.771  & 6.851  & 6.898 & 6.790 \\
Mg$_x$   & 0.5157 & 0.5174 & 0.5131 & 0.511 \\
Mg$_y$   & 0.5603 & 0.5614 & 0.5563 & 0.557 \\
O$_x^1$  & 0.1155 & 0.1128 & 0.1031 & 0.099 \\
O$_y^1$  & 0.4572 & 0.4608 & 0.4654 & 0.464 \\
O$_x^2$  & 0.1914 & 0.1928 & 0.1953 & 0.196 \\
O$_y^2$  & 0.1968 & 0.1995 & 0.2010 & 0.201 \\
O$_z^2$  & 0.5594 & 0.5582 & 0.5510 & 0.561 \\
\end{tabular}
\label{table:struc.ortho}
\end{table}

\begin{table}
\caption{Octahedral rotation angles inferred from cell shape, and mean
Si--O bondlength.  This work gives the most distorted structure, but
also has the smallest cell, supporting the approximation of the effect
of the LDA as an external pressure. [1] Wentzcovitch et al. 1993; [2]
Bukowinski and Wolf 1988; [3] Hemley et al. 1987; [4] Ross and Hazen
1990.
}

\begin{tabular}{rrrl}
    & $\theta(M_2)$ ($^\circ$) & $\phi(R_{25})$ ($^\circ$) & Si--O (\AA)
\\ \tableline
this work &       16.5             &  14.5     &  1.77 \\
other ab initio [1]
          &       15.1             &  13.5     &  1.79 \\
MEG [2] & 10.41    &  7.03     &  1.904\\
SSMEG [3] & 10.8 &  9.8      &  1.776\\
expt, 0~GPa [4] & 14.3 & 11.6      &  1.792 \\
expt, 10.6~GPa [4] & 14.9 & 11.9   &  1.777 \\
\end{tabular}
\label{table:rot.angles}
\end{table}

\begin{table}
\caption{Calculated and measured frequencies of $A_g$ modes of
orthorhombic phase, which are expected to give strong Raman signals.
[1] Navrotsky and Weidner 1989.
}
\begin{tabular}{rl}
this work &  experimental \\
$A_g$ modes & Raman modes [1] \\
\tableline
25.5      &        \\
20.0      &    20.7 \\
18.8      &         \\
15.2      &    15.0 \\
11.4      &    11.4 \\
10.0      &      \\
 8.6      &     8.49 \\
          &     8.34 \\
\end{tabular}
\label{table:raman.modes}
\end{table}

\widetext
\begin{table}
\caption{Calculation of the coefficients of phonons frozen into the
orthorhombic distorted phase.  The phonons are identified by the point
of the BZ at which they occur, with degeneracies given in brackets.  The
depth $E_0$ of the minima of the well has been calculated for unstable
phonons, using $c_0$ and the frequency, assuming no coupling.  Stable
phonons correspond to a well with only one minimum, at the origin, so
$E_0$ is zero.}
\begin{tabular}{ccccl}
    & frequency & $c_0$         & $E_0$ & description \\
    & (THz)& ($\sqrt{\mbox{a.m.u.}}$\,\mbox{\AA}) & (eV) & \\ \hline
$R$ (2)   & 11.8$i$ &  4.58    & 2.99 & rotation of octahedra about $xy$\\
$M$ (1)  & 11.1$i$ &  4.35    & 2.38 & rotation of octahedra about $z$ \\
$X$ (2)  & 4.73$i$ &  2.29    & 0.120 & mostly Mg displacement \\
$R$ (2)  & 3.48$i$ &  0.595   & 0.0021 & mostly Mg displacement \\
$X$ (2)  & 10.5    &  0.551   & stable & Mg and Oi displacement\\
$M$ (1)  & 19.6    &  0.210   & stable & octahedral squash \\
\end{tabular}
\label{tab.depths}
\end{table}
\narrowtext

\begin{table}
\caption{Normal mode coordinates of cubic rotational phonons in the
fully relaxed orthorhombic phase, $c_0$(orth), ionic relaxation in a
cubic cell, $c_0$(cubic), and of the equilibrium structure when the
phonon is frozen in alone, $c'$.  In both orthorhombic cases, the
structure is not as distorted as the minima of the isolated phonons.
Only the magnitudes of the coordinates are given.}
\begin{tabular}{clll}
phonon         & $c'$            &  $c_0$(orth)   & $c_0$(cubic) \\
 \tableline
$M_2$          & 5.45 $\pm$ 0.02 & 4.35           & 4.14 \\
$M_2$ + strain & 5.10 $\pm$ 0.3  & 4.35           & 4.14 \\
$R_{25}$       & 4.7 $\pm$ 0.3   & 4.58           & 4.37 \\
\end{tabular}
\label{tab.frozen}
\end{table}

\begin{figure}
\caption{Cubic, tetragonal and orthorhombic structures of MgSiO$_3$
perovskite.  Twenty atoms of each structure are drawn, corresponding to
four unit cells of the cubic phase and one of the orthorhombic.  Silicon
ions are enclosed in an octahedral cage of oxygen atoms, with magnesium
ions in interstices.  The cell drawn has axes $x$,$y$ and $z$ as
referred to in the text.  The tetragonal phase is formed from the cubic
by rotating the octahedra around the $y$ axis and displacing the
magnesiums along $x$.  In the orthorhombic phase, the oxygen octahedra
are rotated from their cubic positions around $y$ and $z$, and the
magnesium atoms are displaced along $x$ and $y$.  }
\label{fig:str.ortho}
\end{figure}

\begin{figure}
\caption{Charge density through a plane containing Si ion in centre and
two oxygen atoms at edge, showing the large degree of ionicity.}
\label{fig:charge}
\end{figure}

\begin{figure}
\caption{Cubic, tetragonal and orthorhombic phonons at each part of the
BZ of the corresponding cubic phonon.  There are unstable phonons at all
parts of the Brillouin Zone in the cubic phase, but all the phonons in
the orthorhombic phase are stable.  The tetragonal structure results
from freezing in the $R$ rotation phonon to equilibrium, and has only
two unstable modes.  The horizontal axes follow the progression from
cubic to orthorhombic; the arrows link phonons with similar eigenvectors
as the structure becomes more distorted. Loss of degeneracy is due to
the choice of supercell.}
 \label{ortho.phonons}
\end{figure}

\begin{figure}
\caption{Variation of frequency of cubic zone-centre phonons with cell size,
to give Gr\"{u}neisen parameters. }
\label{fig:grunei}
\end{figure}

\begin{figure}
\caption{Double well model of energy as a function of phonon
displacement as described by $E(c) = -\alpha c^2 + \beta c^4$.  The
cubic phase corresponds to the central maximum and the stable
orthorhombic phase corresponds to one of the minima.  The saddle points
correspond to the tetragonal phases: that of the $R_{25}$ rotation has
the lowest energy and is therefore considered as our intermediate phase.}
\label{fig:model.well}
\end{figure}

\begin{figure}
\caption{(a) Energies, (b) forces and (c) stresses for $M_2$ rotation of
the octahedra through different amounts (labelled with change in
fractional coordinate of oxygen).  The energies are fitted to a quartic
well, and the forces to a cubic.  The results with constant cell size ($M_2$
only) are very similar to those with constant octahedral size ($M_2$ +
strain), except for the stresses. These results suggest that the optimum
combination lies somewhere between the two.}
\label{fig.coupl.mstrain}
\end{figure}

\begin{figure}
\caption{One-dimensional example of double well local potentials coupled
by harmonic interactions (the `$\phi^4$' model), which is used to model
the structural phase transitions of \mgsio. A high-symmetry ordered
phase corresponds to all $x_i=0$, and ferro- or antiferromagnetic
distorted phases, with all $x_i$ near the bottom of local wells,
minimise the energy, depending on the sign of the $J_{ij}$. The
harmonic phonon frequencies for oscillation around the central maxima
are used to fit the parameters of this model.  }
\label{fig:model}
\end{figure}

\begin{figure}
\caption{Comparison of the calculated phase boundary for the
orthorhombic--tetragonal phase transition and models of lower mantle
geotherms.  The phase boundary intersects some geotherm models near the
top of the lower mantle, suggesting that the tetragonal phase may be
present in the mantle.  The melting temperature is also shown: this
limits the temperatures possible in the lower mantle, but itself has
considerable error.}
\label{fig:phase.bdy}
\end{figure}

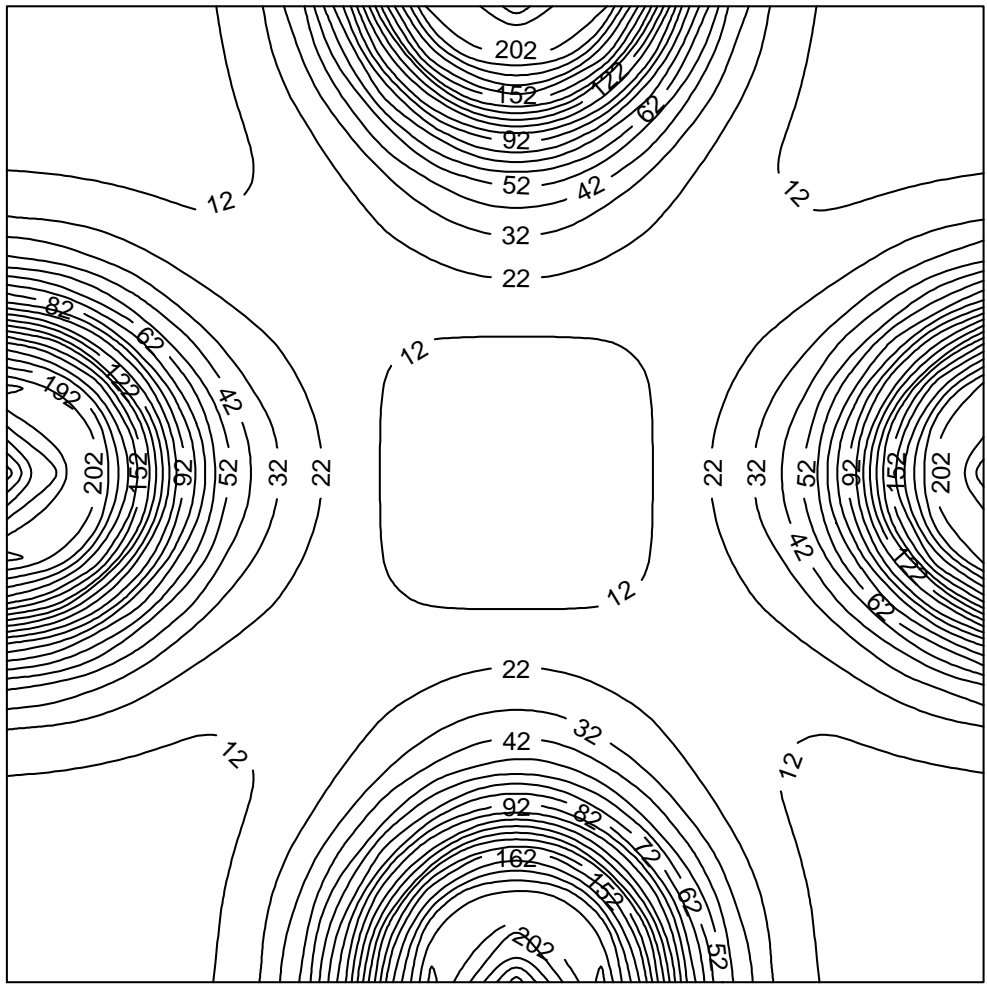
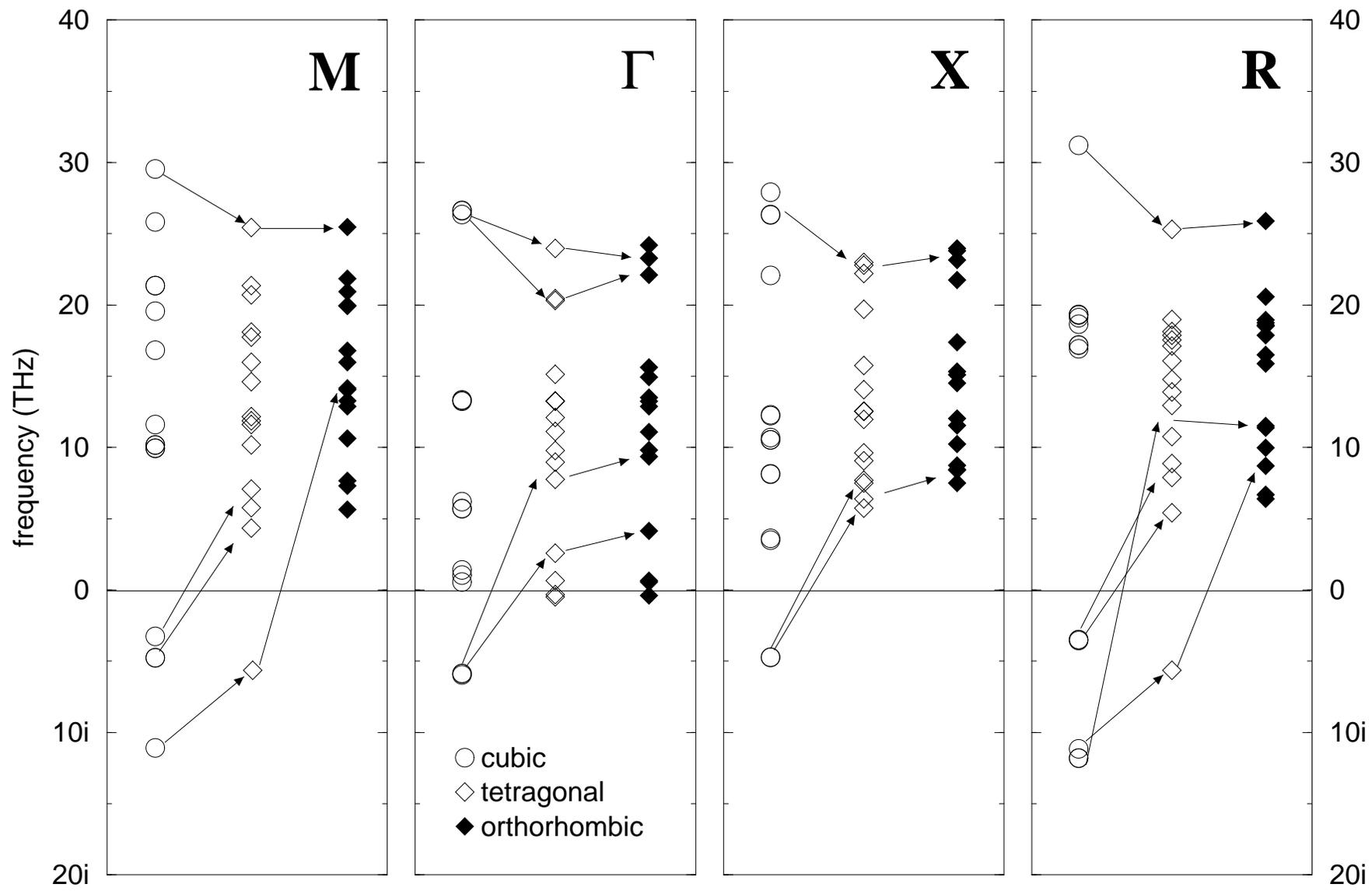
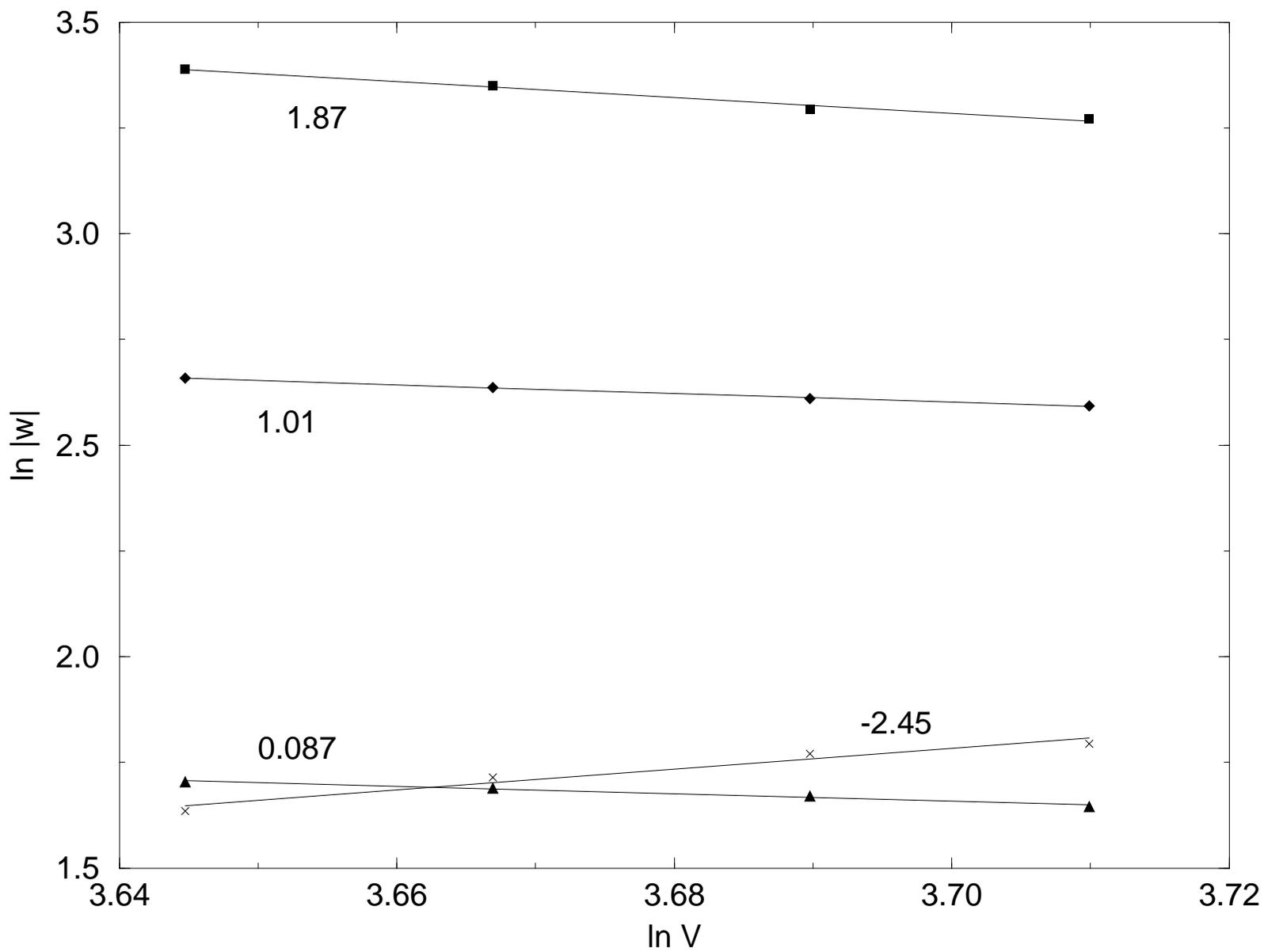
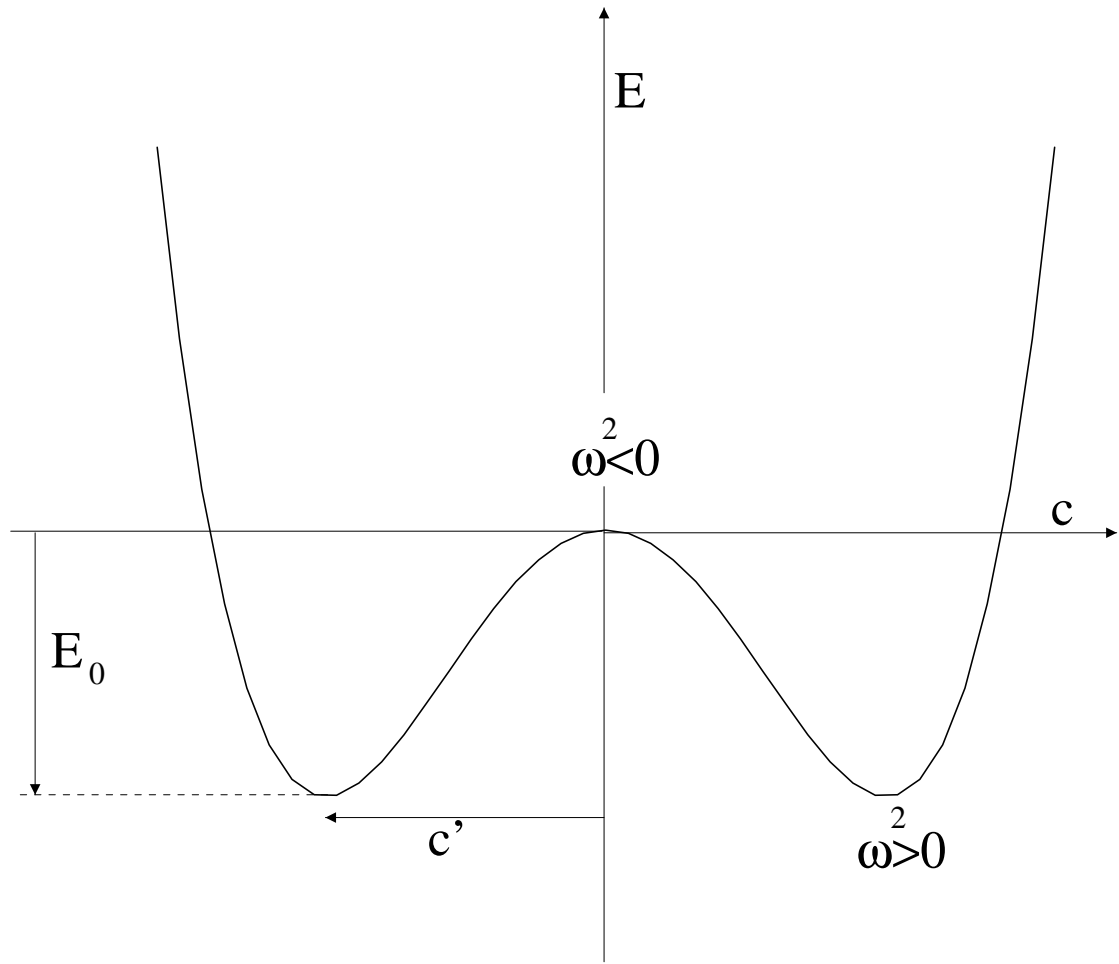
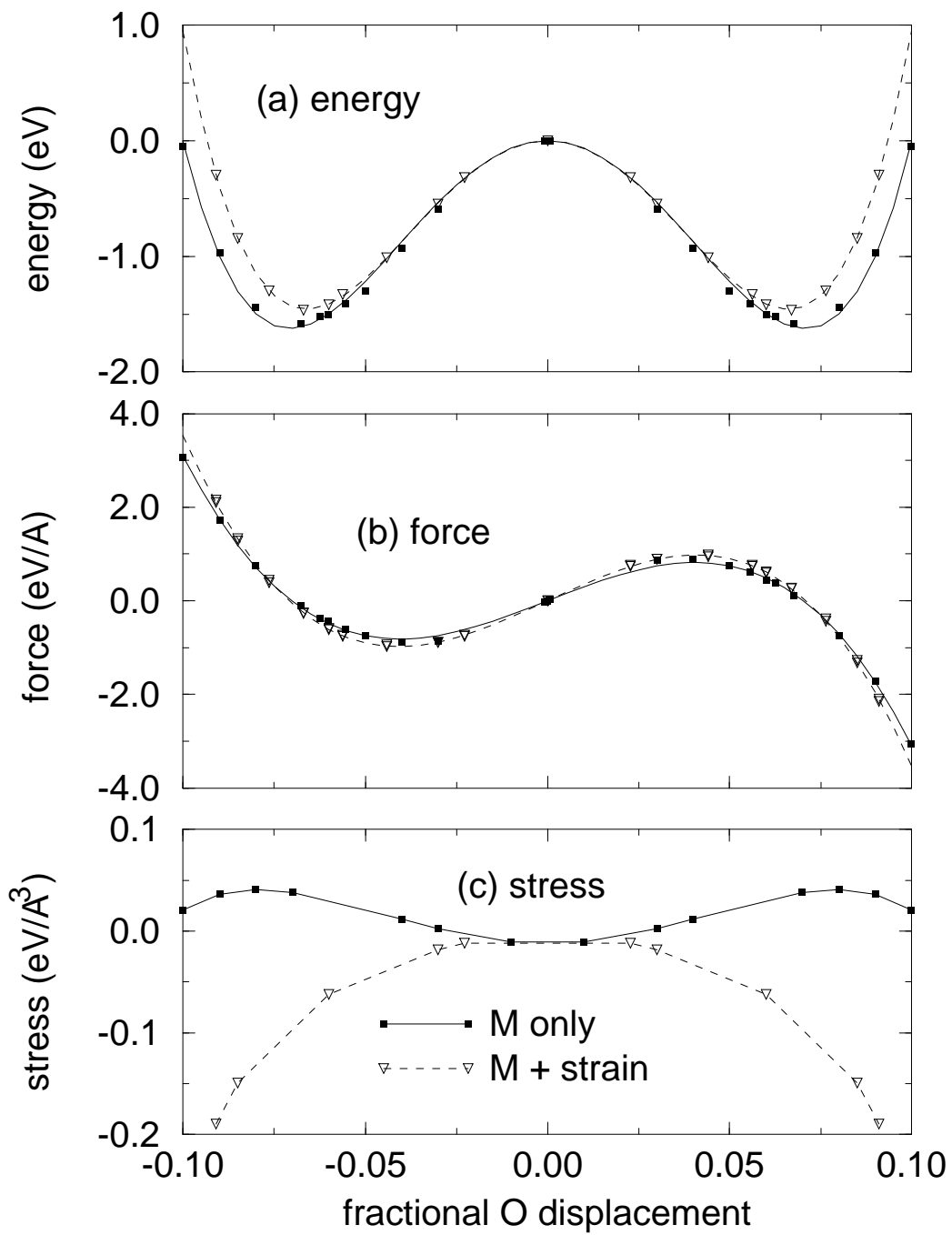
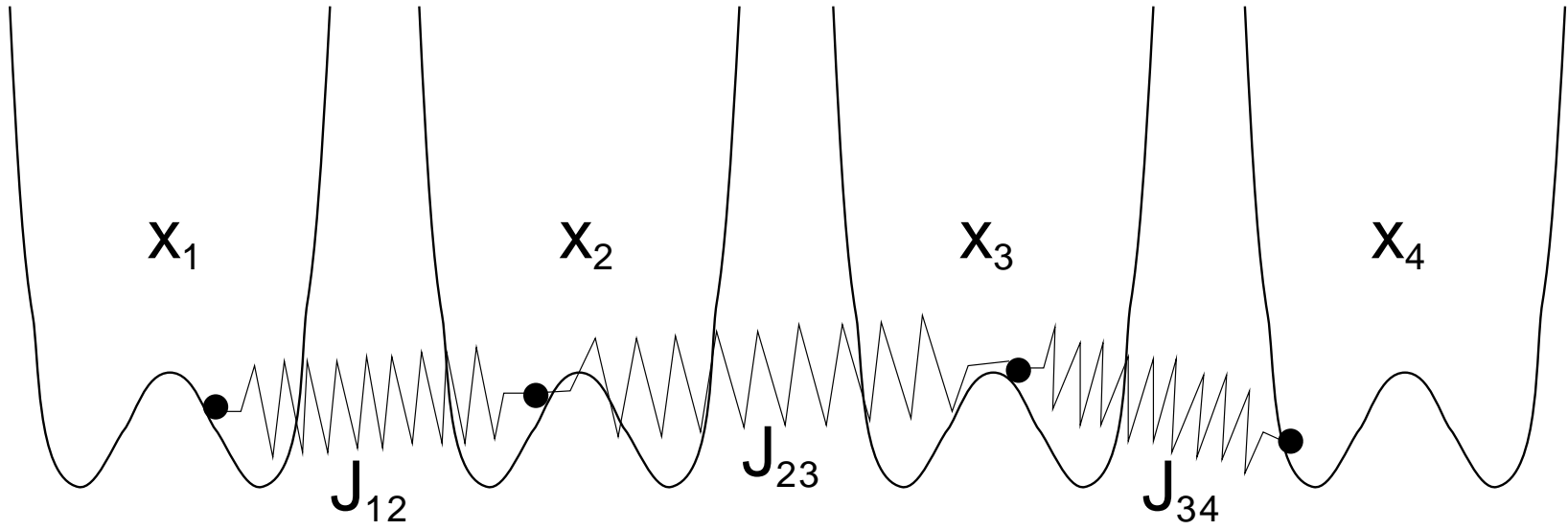
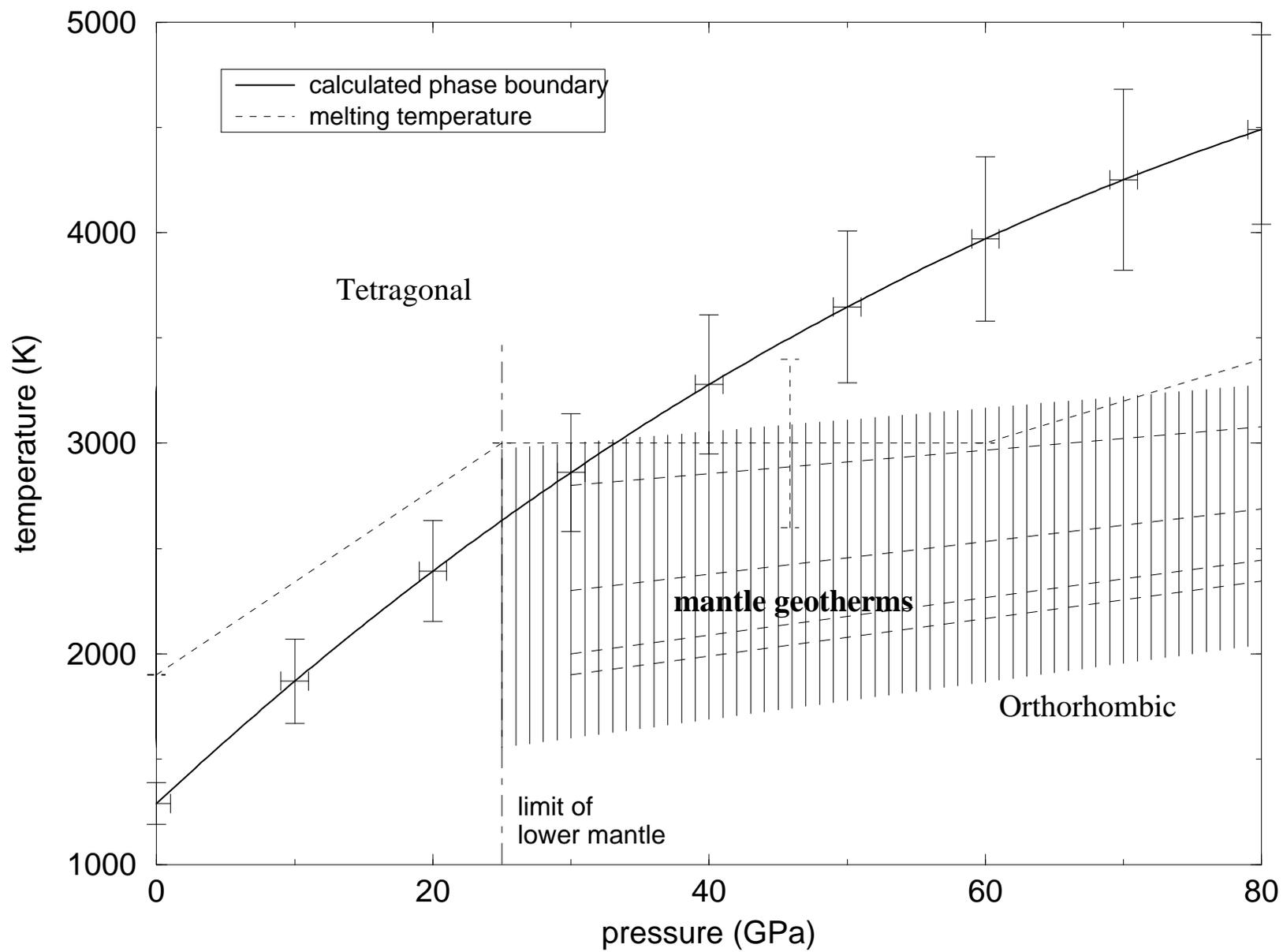
\end{document}